\documentclass[prb,preprint,showpacs,preprintnumbers,amsmath,amssymb,superscriptaddress]{revtex4-1} 
\usepackage{graphicx}
\usepackage{dcolumn}
\usepackage{bm}
\usepackage{setspace}
\usepackage{subfigure}
\usepackage{color}

\begin{document}

\preprint{APS/123-QED}

\title{Theoretical investigation of the evolution of the topological phase of Bi$_{2}$Se$_{3}$ under mechanical strain}

\author{Steve M. Young}
\author{Sugata Chowdhury}
\affiliation{
 The Makineni Theoretical Laboratories, Department of Chemistry, University of Pennsylvania, Philadelphia, PA 19104-6323, USA.
 }
\author{Eric J. Walter}
\affiliation{Department of Physics, College of William and Mary, Williamsburg, VA 23187-8795, USA.
}
\author{Eugene J. Mele}
\author{Charles L. Kane}
\affiliation{
 Department of Physics, University of Pennsylvania,  Philadelphia, PA 19104-6323, USA.
}
\author{Andrew M. Rappe}\email{rappe@sas.upenn.edu}
\affiliation{
 The Makineni Theoretical Laboratories, Department of Chemistry, University of Pennsylvania, Philadelphia, PA 19104-6323, USA.
 } 
\date{\today}


\begin{abstract}
 The topological insulating phase results from inversion of the band gap due to spin-orbit coupling at an odd number of time-reversal symmetric points.  In Bi$_2$Se$_3$, this inversion occurs at the $\Gamma$ point.  For bulk Bi$_2$Se$_3$, we have analyzed the effect of arbitrary strain on the $\Gamma$ point band gap using Density Functional Theory.  By computing the band structure both with and without spin-orbit interactions, we consider the effects of strain on the gap via Coulombic interaction and spin-orbit interaction separately. While compressive strain acts to decrease the Coulombic gap, it also increases the strength of the spin-orbit interaction, increasing the inverted gap.  Comparison with Bi$_2$Te$_3$ supports the conclusion that effects on both Coulombic and spin-orbit interactions are critical to understanding the behavior of topological insulators under strain, and we propose that the topological insulating phase can be effectively manipulated by inducing strain through chemical substitution.      
\end{abstract}

\maketitle

\clearpage

\section{Introduction}
The discovery of the topological insulating (TI) phase of materials has garnered intense interest from the condensed matter physics community and spawned numerous investigations to explore the nature of its origins and effects.~\cite{Kane05p226801,Kane05p146802,Fu07p106803,Moore07p121306,Roy09p195322,Hasan10p3045}  These materials have an insulating gap in the bulk, while also possessing conducting, gapless edge or surface states that are protected by time-reversal symmetry. The prediction and observation of this phase have been made in real materials,~\cite{Bernevig06p1757,Fu07p045302,Konig07p766,Hsieh08p970,Chen09p178,Xia09p398,Zhang09p438} reinforcing its status as a topic of interest and importance.  While most investigations have focused on fundamental physical properties, the unique properties of the topological insulating phase suggest several practical applications, including spintronics and quantum computation.~\cite{Fu08p096407,Qi08p195424,Qi09p1184,Essin09p146805} However, significant progress towards technological applications will require deep understanding of the dependence of the fundamental physics on material structure and composition.  
In this work, we investigate the relationship between the topological insulating phase and the elastic properties of bismuth selenide.  Bi$_{2}$Se$_{3}$ is chemically stable, easy to synthesize, and exhibits a robust topological phase. Combined with the existing theoretical and experimental studies,~\cite{Chen09p178,Xia09p398,Mooser56p492,Black57p240,Zhang09p053114,Peng10p225,Li09p5054,Zhang10p584,Zhang09p266803} it has emerged as the prototypical topological insulator, and is a natural choice for a preliminary investigation. There has been some interest in the effect of mechanical strain on topological effects,~\cite{Zhang11p24,Brune11p2627} but to our knowledge no systematic investigation has been performed.  Here we use {\em ab initio} methods to evaluate the elastic properties of Bi$_{2}$Se$_{3}$, and to connect these to the properties of the topological insulating phase.  
\begin{figure}[h]
\centering
\includegraphics[height=2.6in,width=3.0in]{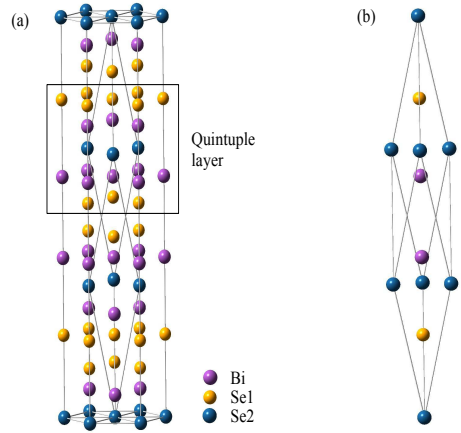}
\caption{{\bf{Crystal Structure of Bi$_{2}$Se$_{3}$ }}: (a) The rhombohedral crystal structure of Bi$_{2}$Se$_{3}$ consists of hexagonal planes of Bi and Se stacked on top of each other along the $z$-direction. A quintuple layer with Se1-Bi-Se2-Bi-Se1 is indicated by the black square, where (1) and (2) refer to different lattice positions. (b) Rhombohedral unit cell of Bi$_{2}$Se$_{3}$.}
\label{fig:Crystal_Structure}
\end{figure}

\section{METHODOLOGY}
Density functional theory calculations were performed using the Perdew-Burke-Ernzerhof-type~\cite{Perdew96p3865} generalized gradient approximation (GGA) implemented in the Quantum-Espresso~\cite{Giannozzi09p395502} code. All atoms are represented by norm-conserving, optimized, designed nonlocal~\cite{Ramer99p12471,Rappe90p1227} pseudopotentials generated using the OPIUM package.~\cite{opium} Pseudopotentials were constructed with and without spin-orbit coupling. All calculations are performed with a plane-wave cutoff of 50~Ry on an $8\times8\times8$ Monkhorst-Pack~\cite{Monkhorst76p5188} k-point mesh. The lattice parameters were taken from experiments  ({\em{a}} = 4.138~\AA~and {\em{c}} = 28.64~\AA).~\cite{Wiese60p13} We calculated band structures along the high symmetry lines in the Brillouin-zone (BZ). 
After fixing the lattice parameters to their experimental values, the atomic coordinates were relaxed to generate the reference structure.  It should be noted that the calculations give significant, nonzero stress for the crystal in this geometry.

Various strains were applied relative to the reference structure, including positive and negative uniaxial and shear strains up to 2\%, as well as several combinations thereof.  The atomic lattice coordinates were relaxed for each strain configuration, and the total and band gap energies were computed both with and without spin-orbit coupling. 
Multiple regression analysis was performed to find the linear and quadratic dependence of the energy on strain tensor components.  This yielded the elastic stiffness and stress tensors. 

It is known that in bismuth selenide the topological index distinguishing ordinary insulating from topological insulating behavior is controlled by band inversion at the $\Gamma$ point. 
Thus, a band gap stress~$\sigma^\Gamma$ and band gap stiffness~$c^\Gamma$ were defined as the linear and quadratic coefficients relating the $\Gamma$ point band gap to strain, by the same procudure used to determine the elastic tensors.   
\begin{eqnarray}
 \Delta E_g^\Gamma(\epsilon) &= \frac{1}{2} c^\Gamma_{ijkl}\epsilon_{ij}\epsilon_{kl} + \sigma^\Gamma_{ij}(0)\epsilon_{ij} 
\end{eqnarray}
In both cases, only the tensor elements unique under the symmetry operations of the space group of bismuth selenide (R$\bar{3}$m, No. 166) were allowed as degrees of freeedom.  The stiffness and stress tensors, in Voigt notation, must have the forms
\begin{eqnarray}
c=\left[ \begin{array}{c  c  c   c   c   c}
      c_{11} & c_{12} & c_{13} & c_{14}  & 0              & 0       \\
      c_{12} & c_{11} & c_{13} &-c_{14}  & 0              & 0       \\
      c_{13} & c_{13} & c_{33} & 0       & 0              & 0       \\
      c_{14} &-c_{14} & 0      & c_{44}  & 0              & 0       \\
      0      & 0      & 0      & 0       & c_{44}         & c_{14}  \\
      0      & 0      & 0      & 0       & c_{14}         & c_{66}
         \end{array} \right] , \qquad 
\sigma=\left[ \begin{array}{c }
      \sigma_{1} \\
      \sigma_{1} \\
      \sigma_{3} \\
        0            \\
        0            \\
        0        
         \end{array} \right]
\end{eqnarray}

\section{RESULTS AND DISCUSSION}
The bulk crystal structure of Bi$_{2}$Se$_{3}$ is rhombohedral with space group D$^{5}_{3d}$ (R$\bar{3}$m, No. 166) \cite{Wiese60p13}, shown in Fig. 1. The primitive unit cell has two Bi and three Se atoms, and the atomic plane arrangement is Se(1)-Bi-Se(2)-Bi-Se(1), where Se(1) and Se(2) indicate the two different types of selenium atom in the crystal. In the hexagonal supercell, the structure can be described as quintuple layers (QL) (square region in Fig.~\ref{fig:Crystal_Structure} of atoms stacked along the trigonal axis (three-fold rotational axis). 

The band structure of Bi$_{2}$Se$_{3}$ and related compounds have been theoretically predicted~\cite{Xia09p398,Zhang09p438,Mishra97p461,Zhang10p065013} and 
\begin{figure}[h]
\begin{center}
  \subfigure[NSO]{\label{fig:SRL}\includegraphics[scale=0.65]{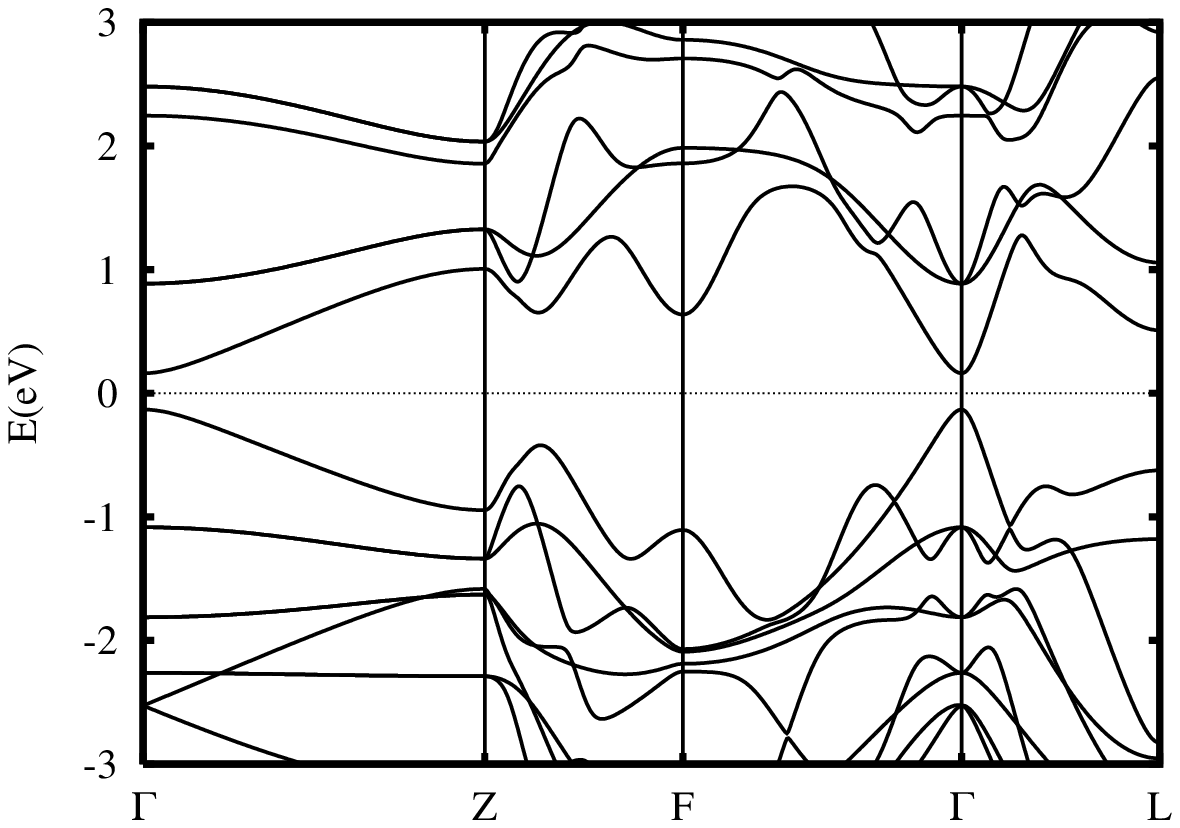}}                
  \subfigure[SO]{\label{fig:FRL}\includegraphics[scale=0.65]{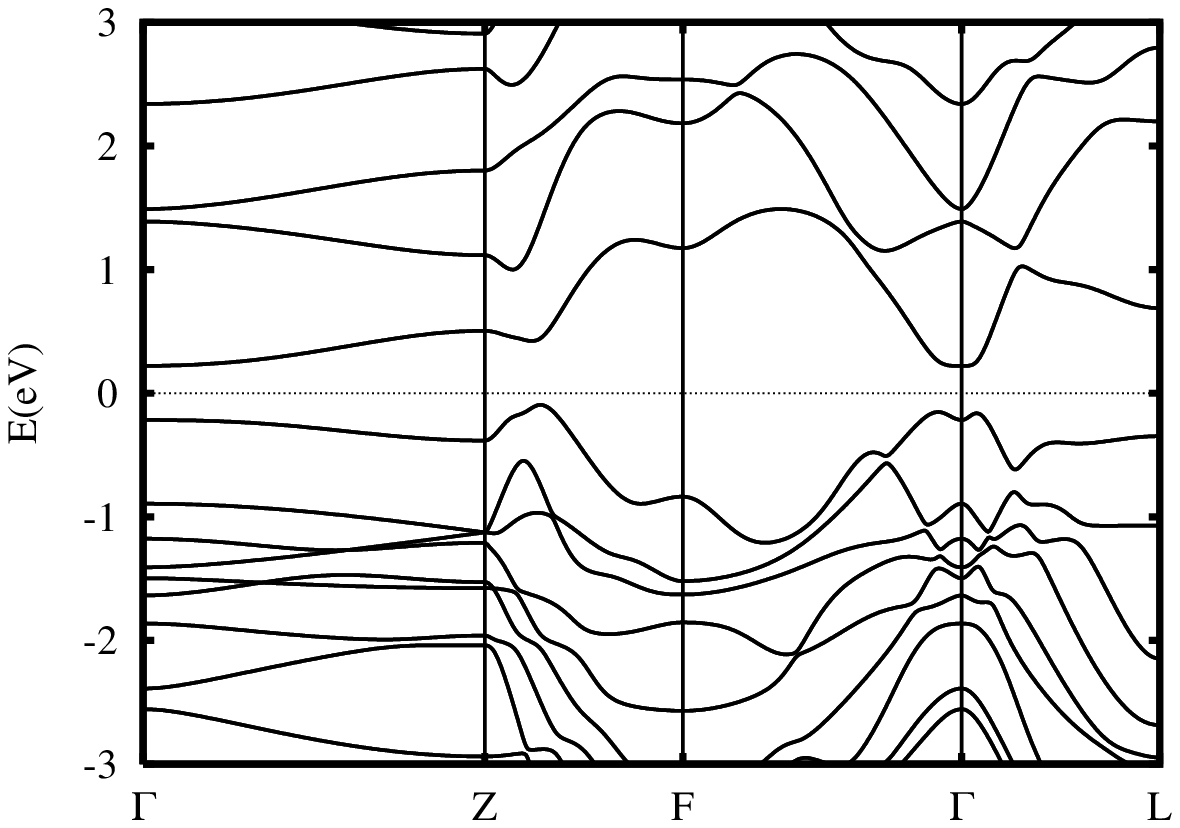}}
  \caption{Band structure in the reference strain state of of Bi$_{2}$Se$_{3}$ (a) excluding (NSO) and (b) including (SO) spin-orbit effects. The dashed line indicates the Fermi level.}
  \label{fig:Band_Structure_Exp}
\end{center}
\end{figure}
\begin{table}[h]
  \begin{center}
  \begin{tabular*}{\columnwidth}{lr}
  \hline
    Element  \qquad \qquad & Coefficient (GPa)\\ 
  \hline
   $c_{11}$             &   91.8   $\pm$   1.0 \\  
   $c_{33}$             &   57.4   $\pm$   1.4 \\
   $c_{44}$             &   45.8  $\pm$   1.0  \\
   $c_{66}$             &   56.2  $\pm$   1.8  \\
   $c_{12}$             &   36.6  $\pm$    1.2 \\
   $c_{13}$             &   38.6  $\pm$    2.0 \\
   $c_{14}$             &   24.2  $\pm$    1.8 \\ \hline
   $\sigma_1$            &  -3.447  $\pm$   0.007 \\ 
   $\sigma_3$            &  -1.977  $\pm$  0.010 \\ \hline
  \end{tabular*}
  \caption{The unique elements of the elastic stiffness and stress tensors of Bi$_{2}$Se$_{3}$.  Spin-orbit coupling has been included.}
  \label{tab:estiff}
  \end{center}
\end{table}
experimentally observed.~\cite{Hsieh08p970,Chen09p178}
In our calculations, the band gap of the unrelaxed, experimental structure of Bi$_{2}$Se$_{3}$ is 0.3 eV, which is consistent with the experimental data and other calculations.~\cite{Hsieh08p970,Mooser56p492,Black57p240} Fig.~\ref{fig:Band_Structure_Exp} shows the band structure with and without spin-orbit interaction, and both are in excellent agreement with previous results.~\cite{Zhang09p438} 
\begin{table}[h]
\centering
\begin{tabular*}{\columnwidth}{ l c c }
\hline
	             & NSO                      &  SO  \\
Element  \qquad  \qquad    &         Coefficient (eV)   \qquad  \qquad  &   Coefficient (eV)                \\ \hline
$c^\Gamma_{11}$                  &   -67.8    $\pm$   2.6   &   35.4   $\pm$    5.6 \\  
$c^\Gamma_{33}$                  &    55.4    $\pm$   3.6   &  -60.1   $\pm$    7.6 \\
$c^\Gamma_{44}$                  &   -58.0    $\pm$   2.6   &   23.4   $\pm$   5.6                       \\
$c^\Gamma_{66}$                  &   -126.6  $\pm$   4.4   &   69.4   $\pm$   9.6 \\
$c^\Gamma_{12}$                  &    60.2    $\pm$   3.2   &  -33.8   $\pm$    6.8 \\
$c^\Gamma_{13}$                  &     6.4     $\pm$   4.2   &  -12.6   $\pm$     10.8 \\
$c^\Gamma_{14}$                  &   -70.0    $\pm$   4.2   &   45.6   $\pm$      9.0  \\ \hline
$\sigma^\Gamma_1$                &   0.16  $\pm$   0.017 &  -1.67  $\pm$   0.037 \\ 
$\sigma^\Gamma_3$                &    4.33  $\pm$   0.023 &  -5.27  $\pm$  0.051 \\ \hline
\end{tabular*}
\caption{The unique elements of the $\Gamma$ band gap stiffness and stress tensors  excluding (NSO) or including (SO) spin-orbit coupling}
\label{tab:gapstiff}
\end{table}
\begin{figure}[h]
\begin{center}
\includegraphics[height=4in,width=5in]{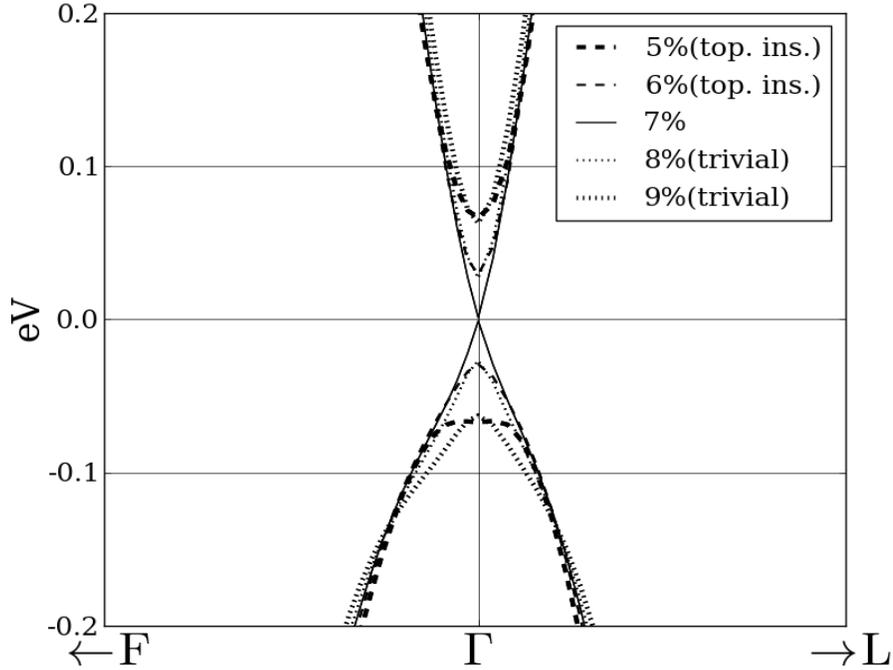}
\caption{Band structure of Bi$_{2}$Se$_{3}$ near the $\Gamma$ point as $<$111$>$ uniaxial strain from 5\% to 9\% drives the topological phase transition.}
\label{fig:cross}
\end{center}
\end{figure}
The computed elastic and band gap tensor components are given in Table~\ref{tab:estiff} and Table~\ref{tab:gapstiff}, respectively. 
First, we note that the gap stress shown in Table~\ref{tab:gapstiff} for strain normal to the plane of the quintuple layers($\sigma^\Gamma_3$) is much higher than for strain in the plane($\sigma^\Gamma_1$), which is consistent with the notion that inter-layer interactions are more important in determining the band gap than intra-layer interactions.  
Second, we observe the change in sign of $\sigma^\Gamma$ when spin-orbit interactions are turned off or on. This is to be expected because the spin orbit interaction leads to an inversion of the conduction and valence bands.
Thus, strain reduces the bandgap for the trivial (un-inverted) phase and increases the gap for the topological (inverted) phase.
Third, the magnitude of the gap stress is larger when spin-orbit interactions are present.  From a tight-binding perspective, compressive strain not only strengthens the Coulombic interaction between sites, increasing the associated hopping coefficient and reducing the conventional gap, but also magnifies the spin-orbit effect and its hopping coefficient, increasing the topological gap.  Thus, comparing the gap stress with and without spin-orbit interactions provides some insight into the effects of strain on the essential physics of the system.        

Using the above tensors, we can predict that the topological phase transition will occur at 6.4\% uniaxial strain in the $<$111$>$ direction.  In Fig.~\ref{fig:cross}, the onset of the topological insulating phase at 7\% strain can be observed through changes in the band structure as strain increases.  At the transition point, the Dirac cone characteristic of the phase transition is distinctly observable.  Of course, such strains are difficult to achieve experimentally.  According to the computed elastic tensors, around 2~GPa of uniaxial tensile stress would be required to drive the phase transition, well past the yield stress.  However, large strains may be possible by introducing internal stress through  chemical substitution.\\
\begin{table}[h]
\centering
\begin{tabular*}{\columnwidth}{ l | c | c | c}
\hline
                                &        Bi$_{2}$Se$_{3}$(reference) &   Bi$_{2}$Se$_{3}$(strained) &  Bi$_{2}$Te$_{3}$(reference) \\
\hline
Lattice parameters(\AA)&        {\em{a}} = 4.138, {\em{c}} = 28.64& {\em{a}} = 4.358, {\em{c}} = 30.46 & {\em{a}} = 4.358, {\em{c}} = 30.46\\
\hline
Anion radius         (\AA)      &    ~1.98        & ~1.98         &        ~2.21  \\
\hline
NSO Gap  (eV)     &     ~0.02     &        ~0.31            &       ~0.20\\
SO Gap      (eV)  &     ~0.42     &        -0.06           &       ~0.63\\
\hline
\end{tabular*}
\caption{Comparison of reference Bi$_{2}$Se$_{3}$,  Bi$_{2}$Se$_{3}$ strained to match reference Bi$_{2}$Se$_{3}$ lattice, and reference Bi$_{2}$Te$_{3}$. Spin-Orbit(SO) and Non-Spin-Orbit(NSO) $\Gamma$ point gaps for all three structures are calculated.}
\label{tab:bt}
\end{table}
Bi$_{2}$Te$_{3}$ is a very similar compound to Bi$_{2}$Se$_{3}$, differing only in substitution of the larger tellurium in place
of selenium, which increases the size of the lattice by about 6\%.  It is also a topological insulator with band inversion occurring at the $\Gamma$ point.  Given the similarity, one may ask if it is reasonable to view Bi$_{2}$Te$_{3}$
as intrinsically strained  Bi$_{2}$Se$_{3}$.  To test this hypothesis, we performed a comparison of Bi$_{2}$Te$_{3}$ to  strained Bi$_{2}$Se$_{3}$.

In order to generate an appropriate reference structure for comparison, Bi$_{2}$Te$_{3}$ was relaxed under identical external stress as the reference Bi$_{2}$Se$_{3}$(this is labeled ``Bi$2$Te$3$(reference)").  The results are shown in Table~\ref{tab:bt}.  
The computed lattice parameters of Bi$_{2}$Te$_{3}$ are in good agreement with experiment.  
The Bi$_{2}$Se$_{3}$ lattice was then strained so that the lattice parameters match that of reference Bi$_{2}$Te$_{3}$(this is labeled ``Bi$2$Se$3$(strained)"). Using the gap stiffness and stress tensors, the band gap of this strained Bi$_{2}$Se$_{3}$ was calculated and compared to the computed band gap of Bi$_{2}$Te$_{3}$.  Without spin-orbit interaction, the strained Bi$_{2}$Se$_{3}$ band gap is similar to the band gap of reference Bi$_{2}$Te$_{3}$.  However, with spin-orbit interaction turned on, the gaps are dramatically different: in the strained Bi$_{2}$Se$_{3}$, the topological gap closes, but in Bi$_{2}$Te$_{3}$ the topological gap is quite large.  This suggests that spin-orbit effects are strongly dependent on the  chemical identity of the anion, and that treating  Bi$_{2}$Te$_{3}$ as strained Bi$_{2}$Se$_{3}$ fails to capture the essential physics.

\section{CONCLUSION}
	We have shown that strain is an important parameter for influencing the topological insulating phase.  For bismuth selenide, the direct band gap at the gamma point, where band inversion occurs, responds to elastic deformation in a way that can be described by adapting the formalism of continuum mechanics.     The critical strain at which the topological phase transition occurs was predicted using the derived band gap stress and stiffness tensors and observed in the computed band structure as a Dirac cone.  
	While it may be possible to tune the band gap with external stress, more interesting is the potential for inducing strain via chemical substitution. Viewing bismuth telluride as chemically strained bismuth selenide, however, fails dramatically, hinting at a complex relationship between chemical composition, material structure, and the physics underlying the topological insulating phase.
 
\section{ACKNOWLEDGMENT}
SMY was supported by the Department of Energy Office of Basic Energy
Sciences, under Grant No. DE-FG02-07ER46431.
SC acknowledges the support of the National Science Foundation,
through the MRSEC program, Grant No. DMR05-20020, as well as computational support from the Center for Piezoelectrics by Design.
AMR acknowledges the support of the Office of Naval Research, under
Grant No. N-000014-00-1-0372, as well as computational support from
the HPCMO.  C.L.K. acknowledges support from NSF grant DMR-0906175. 
EJM was supported by the Department of Energy under Grant No. DE-FG02-ER45118.

\end{document}